\documentclass[a4paper]{article}
\usepackage[utf8]{inputenc}
\DeclareUnicodeCharacter{2212}{-}

\usepackage[english]{babel}

\usepackage{geometry}
\usepackage{pdflscape}

\usepackage{fancyhdr}

\usepackage{lineno}

\usepackage{titlesec}

\usepackage{hyperref}

\usepackage{float}

\usepackage[version=4]{mhchem}

\usepackage{textgreek}

\usepackage{textcomp}

\usepackage{enumitem}

\usepackage{amsmath} 

\usepackage{mathtools}

\usepackage{MnSymbol}

\usepackage{xfrac}

\usepackage{tabu}
\usepackage{booktabs}
\usepackage{dcolumn}
\usepackage{multirow}

\usepackage[para,online,flushleft]{threeparttable}

\usepackage{booktabs}
\usepackage{array}

\usepackage{rotating}

\usepackage{adjustbox} 

\usepackage{graphicx}

\usepackage{caption}

\usepackage{color}

\usepackage{soul} 

\usepackage[labelfont=bf]{caption}

\definecolor{blueMatlab}{rgb}{0, 0.4470, 0.7410}
\definecolor{redMatlab}{rgb}{0.8500, 0.3250, 0.0980}

\usepackage[numbers, round, comma, compress]{natbib}

\usepackage{authblk}

\usepackage{setspace}

\usepackage{wrapfig}





\title{Determination of Nano-sized Adsorbate Mass in Solution using Mechanical Resonators: Elimination of the so far Inseparable Liquid Contribution}

\author[1]{Antonius Armanious\thanks{armanioa@ethz.ch; fredrik.hook@chalmers.se}\thanks{Present address: Laboratory of Food and Soft Materials, Department of Health Sciences and Technology, ETH Z\"urich, Z\"urich, Switzerland}}
\author[1]{Bj\"orn Agnarsson}
\author[2]{Anders Lundgren}
\author[1,3]{Vladimir P. Zhdanov}
\author[1]{Fredrik H\"o\"ok$^\text{*}$}

\affil[1]{Division of Nano and Biological Physics, Department of Physics, Chalmers University of Technology, Gothenburg, Sweden}

\affil[2]{Department of Chemistry and Molecular Biology, University of Gothenburg, Gothenburg, Sweden}

\affil[3]{Boreskov Institute of Catalysis, Russian Academy of Sciences, Novosibirsk, Russia}

\date{}

\begin{document}
\maketitle

\begin{wrapfigure}{r}{0.5\textwidth}
  \centering
    \includegraphics[scale=1.0]{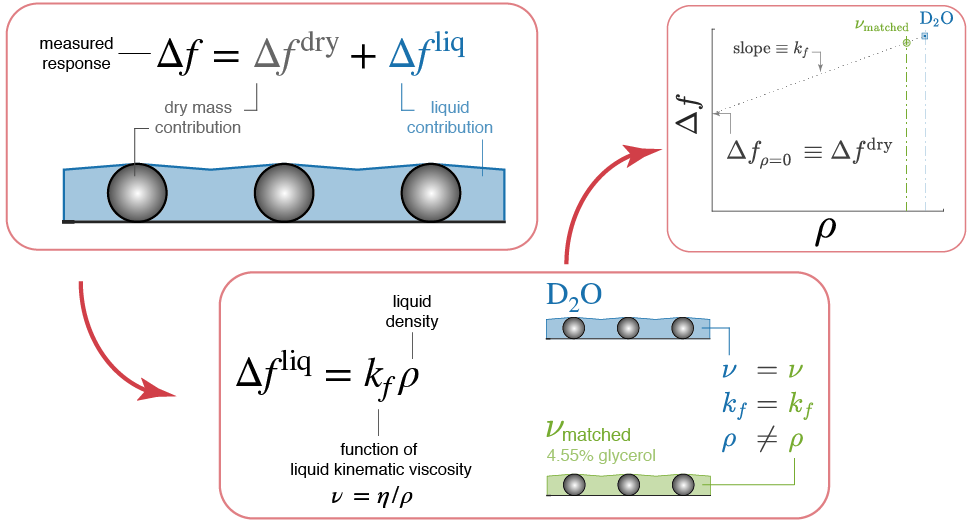}
\end{wrapfigure}

\setcounter{figure}{0}

\textbf{Abstract:} Assumption-free mass quantification of nanofilms, nanoparticles, and (supra)molecular adsorbates in liquid environment remains a key challenge in many branches of science. Mechanical resonators can uniquely determine the mass of essentially any adsorbate; yet, when operating in liquid environment, the liquid dynamically coupled to the adsorbate contributes significantly to the measured response, which complicates data interpretation and impairs quantitative adsorbate mass determination. Employing the Navier-Stokes equation for liquid velocity in contact with an oscillating surface, we show that the liquid contribution can be eliminated by measuring the response in solutions with identical kinematic viscosity but different densities. Guided by this insight, we used quartz crystal microbalance (QCM), one of the most widely-employed mechanical resonator, to demonstrate that kinematic-viscosity matching can be utilized to accurately quantify the dry mass of systems such as adsorbed rigid nanoparticles, tethered biological nanoparticles (lipid vesicles), as well as highly hydrated polymeric films. The same approach applied to the simultaneously measured energy dissipation made it possible to quantify the mechanical properties of the adsorbate and its attachment to the surface, as demonstrated by, for example, probing the hydrodynamic stablization induced by nanoparticles crowding. Finally, we envision that the possibility to simultaneously determine the dry mass and mechanical properties of adsorbates as well as the liquid contributions will provide the experimental tools to use mechanical resonators for applications beyond mass determination, as for example to directly interrogate the orientation, spatial distribution, and binding strength of adsorbates without the need for complementary techniques.


\newpage

\noindent
Determination of the mass of nano-sized adsorbates at solid-air and solid-liquid interfaces is central in many branches of science. Mechanical resonators are unique in their capacity to determine the mass of essentially any type of adsorbate at any surface without \textit{a priori} knowledge about their physicochemical properties. In vacuum and air, this is possible due to a direct proportionality between the measured changes in resonance frequency, $\Delta f$, and the adsorbed mass per unit area, $\Delta m$. For the quartz crystal microbalance (QCM), which is the most widely used mechanical resonator, this relationship is known as the Sauerbrey equation \cite{Sauerbrey1959}: \par
\begin{equation}
\Delta m = -C_\mathrm{Q} \cdot \Delta f_n/n,
\label{eq:Sauerbrey}
\end{equation} \par
\noindent
where $n$ is the overtone number and $C_\mathrm{Q}$ is the mass sensitivity constant, which for a $5 \; \mathrm{MHz}$ crystal is $18 \; \mathrm{ng \cdot Hz^{-1} \cdot cm^{-2}}$ (Ref.\ \citenum{Reviakine2011}).  QCM as well as many other mechanical resonators \cite{Ward1990, Kosaka2014, Martinez-Martin2017} can also operate in liquid environment and be further extended to simultaneously monitor both $\Delta f$ and changes in energy dissipation, $\Delta D$ \cite{Tessler1994, Rodahl1995}. However, when operated in liquid environment, not only the adsorbate, but also the surrounding liquid becomes dynamically coupled to the oscillating crystal \cite{Kanazawa1985}, a feature that is shared by all mechanical resonators operating in liquid. While this can in principle provide unique information about the physicochemical properties of an adsorbate, deconvoluting the entangled information in the measured response is an unresolved analytical challenge, as the measured response in liquid environment is, in a complex manner, dependent on the mass of the adsorbate and the dynamically-coupled liquid. This dependence is manifested through changes in both $\Delta f_n$ and the energy dissipation, $\Delta D_n$, of the resonator \cite{Hook2001}.

Still, the rich information contained in combined $\Delta f_n$ and $\Delta D_n$ measurements, especially when complemented with theoretical models representing the response for viscoelastic films \cite{Wolff1997, Voinova1999, Johannsmann2008a}, has turned out very valuable in multiple research areas, including hydration analysis of organic polymers \cite{Arazoe2016, Liu2018}, proteins \cite{Reinisch1989} and biological membranes  \cite{Reinisch1989, Fu2018}. QCM has also been widely applied to monitor and characterize thin films, including porosity determination \cite{Feng1994, Shpigel2019}, monitoring the growth of mesoporous materials \cite{Walcarius2007}, probing responsiveness of polymeric coatings \cite{Trachsel2020}, and characterization of biomimetic membranes \cite{Richter2007, Shen2014}. Additionally, QCM has been extensively employed to study discrete adsorates such as aboitic and biological macromolecules and nanoparticles (NPs). These studies included investigation of biomolecular interactions of proteins \cite{Dubacheva2017, GutierrezSanchez2015} and viruses \cite{Parveen2019, DiIorio2019}; structure \cite{Mateos-Gil2016, PaulRoach2005}, confirmation \cite{Tsortos2008, Papadakis2010} and orientation changes \cite{Bisker2016} of bio-macromolecules; spatial distribution \cite{Schmudde2016}, size \cite{Tellechea2009, Olsson2013, Gillissen2017a}, deformation \cite{Reviakine2012, Gillissen2017} and dissolution \cite{Zou2021} of NPs; protein corona formation on amyloids \cite{Pilkington2018}; NPs interactions with biomimetic membranes \cite{Melby2017, Lochbaum2021}; as well as bioanalytical sensors development \cite{Janshoff2000a, Patolsky2001, Cooper2001, Webster2014}, 

Yet, the inseparable nature of the adsorbate and liquid contributions to the measured response puts significant limitation on quantitative interpretation of adsorption measurements in liquid environment. In particular, this shortcoming makes mass determination using mechanical resonators non-conclusive, unless complementary methods, such as surface plasmon resonance (SPR), ellipsometry and/or atomic force microscopy (AFM) are employed in parallel \cite{Shpigel2019, Richter2003, ErikReimhult2004a,VanderMeulen2014}.

The use of combined experimental setups, e.g., QCM/SPR or QCM/AFM, comes with many challenges, such as operation at different spatial scales, requirement for very complex setups that are inaccessible for most users, and/or the need of pre-treatment procedures that tend to change the properties of the samples under investigation. Moreover, all complementary methods require accurate knowledge, or determination, of the physical properties of adsorbates, such as refractive index or density. Accordingly, there has in the recent years been an intense, albeit so far unsuccessful search for means to separate the contribution of coupled liquid using QCM measurements alone \cite{Reviakine2011}. Such self-sufficient approach would obviate the need for complementary methods, and provide quantitative information about the adsorbed dry mass and the coupled liquid contribution, as well as information about the dissipative energy loss associated with the adsorbate.

We here address this challenge through insights gained from the general form of the Navier-Stokes equation for the velocity of liquid motion \cite{LANDAU1987}, which for QCM resonators, as well as most mechanical resonators, is reduced to (see Section S1 in the SI for detailed derivation) \par
\begin{equation}
\frac{\partial {\bf v}}{\partial t} = \nu \nabla^2  {\bf v},
\label{eq:NS}
\end{equation} \par
\noindent
where ${\bf v}$ is the liquid velocity, $\nu \equiv \eta /\rho$ is the liquid kinematic viscosity ($\eta$ and $\rho$ are the viscosity and density of the liquid), and $\nabla^2$ represents the Laplacian operator. For firmly bound rigid adsorbates or for adsorbates with appreciably larger inertia than that of the adjacent liquid, the frequency induced by the dry mass of the adsorbate, $\Delta f^\mathrm{dry}$, and the liquid-related change in frequency, $\Delta f^{\rm liq}$, are additive. For such systems, the corresponding no-slip boundary condition at the surface- and adsorbate-liquid interface (i.e., the condition stating that the liquid velocity at the interface is the same as that of the sensor surface and adsorbate) depends only on the geometry of the interface. Accordingly, the solution of Eq.\ (\ref{eq:NS}) depends only on $\nu$ and the interfacial geometry.  Further, the liquid-related change in frequency, $\Delta f^{\rm liq}$, is proportional to the liquid-related force acting at the surface- and adsorbate-liquid interfaces, which in turn is proportional to $\eta$ and the gradient of ${\bf v}$ at the interface along the normal direction \cite{LANDAU1987}. Hence, the spatial derivative of ${\bf v}$ depends on ${\nu}$ and the interfacial geometry only, and accordingly the force integrated over the interface as well as the liquid-related frequency shift, $\Delta f^{\rm liq}$, can be represented as $\eta F(\nu)$, where $F(\nu)$ is a function depending on ${\nu}$ and the interfacial geometry only. Focusing on $\Delta f^{\rm liq}$, and by multiplying and dividing the corresponding expressions by $\rho$, while taking into account that $\eta /\rho = \nu $, we can represent $\Delta f^{\rm liq}$ as $\rho \nu F(\nu)$, i.e., as a product of $\rho$ and another function [$\nu F(\nu)$] which depends on $\nu$ and the geometry only. Using $k_f$ for the designation of the latter function, we have $\Delta f^{\rm liq}=k_f \cdot \rho$, and thus, the measured frequency response, $\Delta f$, is given by \par
\begin{equation}
\Delta f_i=\Delta f^{\rm dry} + \Delta f^{\rm liq}_i
 = \Delta f^{\rm dry} + k_f \cdot \rho_i ,
\label{eq:fmeasured}
\end{equation} \par
\noindent
where $i$ is the subscript introduced in order to specify the type of liquid [compared to Eq.\ (\ref{eq:Sauerbrey}), we do not indicate $n$ here]. For the change in dissipation induced by a firmly bound adsorbate, by analogy, we obtain (see Section S1 in the SI for detailed derivation) \par
\begin{equation}
\Delta D_i=\Delta D^{\rm mech} + \Delta D^{\rm liq}_i
 = \Delta D^{\rm mech} + k_D \cdot \eta_i ,
\label{eq:Dmeasured}
\end{equation} \par
\noindent
where $\Delta D^{\rm mech}$ is the term related to the mechanically dissipated energy, $\Delta D^{\rm liq}_i$ is the liquid-related term, and $k_D$ is the corresponding constant which depends on $\nu_i$ and the interfacial geometry. 

Physically, $\Delta f^\mathrm{dry}$ and $\Delta D^\mathrm{mech}$ are associated with the motion of the adsorbate by itself, i.e., its motion in the absence of the surrounding liquid, with $k_f \cdot \rho_i$ and $k_D \cdot \eta_i$ being the Navier-Stokes-related liquid contributions. Based on the adsorbate system and liquids used, $\Delta f^\mathrm{dry}$  and $\Delta D^\mathrm{mech}$ might also include adsorbed liquid provided these liquids are not described by the Navier-Stokes equation. This distinction will be addressed in more details when discussing the experimental results in the following sections.

If measurements are performed in one liquid alone, Eqs.\ (\ref{eq:fmeasured}) and (\ref{eq:Dmeasured}) are not very helpful for practical applications because the dependence of $k_f$ and $k_D$ on $\nu$ and the interfacial geometry can be analytically expressed only for flat surfaces [e.g., Eq.\ (4) in Ref.\ \citenum{Kanazawa1985}]. In principle, for all other cases, this dependence can be calculated numerically \cite{Johannsmann2009} or analytically. Numerical estimations are, however, neither straightforward nor universal, whereas analytical expressions can practically be derived only by employing severe approximations, e.g., by using Brinkman's mean-field 1D equation as was previously done in order to describe the liquid-related response of the QCM for a rough surface \cite{LeonidDaikhin2002} (Section S1 in the SI).  Our strategy is different and generally applicable. In particular, we suggest to quantify $\Delta f^{\rm dry}$ and $\Delta D^{\rm mech}$ by determining $k_f$ and $k_D$ via measurements in different solutions matched with respect to their kinematic viscosity, i.e., in solutions with identical $\nu_i$ but different $\rho_i$ and $\eta_i$. According to Eqs.\ (\ref{eq:fmeasured}) and (\ref{eq:Dmeasured}), this will allow us to exclude the contribution of coupled liquid to the QCM signal by extrapolating the measured values to $\rho=0$ and $\eta =0$ respectively, and thereby extract $\Delta f^{\rm dry}$ and $\Delta D^{\rm mech}$ from the respective intercepts, without requiring any information about the geometry of the adsorbate. Above, this approach has been derived for rigid adsorbate systems. Our experiments presented below show that it can be applicable and remain fairly accurate even for viscoelastic adsorbate systems, such as biological soft matter.

To experimentally evaluate the applicability of the proposed approach, we first estimated the dry mass of $150 \; \mathrm{nm}$ \ce{SiO2} nanoparticles (NPs) electrostatically adsorbed on a flat QCM surface at coverages, $\theta$, ranging from 1 to 54\%. We also disentangled the liquid-related contribution to $\Delta D$ from the contribution associated with the attachment between the NPs and the surface. The latter was explored to gain new insights into the postulated influence of surface crowding of NPs on the stiffness of the attachment between each NP and the surface  \cite{Johannsmann2009}. We also explored this approach for adsorption of carboxylated $70 \; \mathrm{nm}$ gold (Au) NPs, as well as PEGylated Au NPs, to positively charged flat surfaces. For these experiments, we have varied the stiffness of the linker between the Au NPs and the surface by either varying the ionic strength of the solution or the length of the PEG linker. Then we tested the applicability of the approach on flexible biological NPs by assessing the dry mass of lipid vesicles tethered to supported lipid bilayer (SLB) by DNA linkers. The similarity in structure and properties of lipids vesicles to exosomes and enveloped viruses make them ideal models for assessing the applicability of the approach on biological nanoparticles. Finally, we extended the applicability of the approach beyond NPs by quantifying the dry mass of adsorbed humic acids (HAs), a highly hydrated \cite{Armanious2014} and permeable  \cite{Duval2005} natural polymeric material composed of supramolecular assemblies of relatively small molecules, ranging from $\simeq 200$ to $2500$ Da (Ref.\ \citenum{Sutton2005}) with key environmental importance, due to the role they play in, among others, transport of microbes \cite{Armanious2016}, suppression of the release of green-house gases \cite{Klupfel2014} and photochemical oxidation of pollutants \cite{Latch2006}.

\begin{figure}[H] 
	\begin{center}
	\includegraphics[scale=1.0]{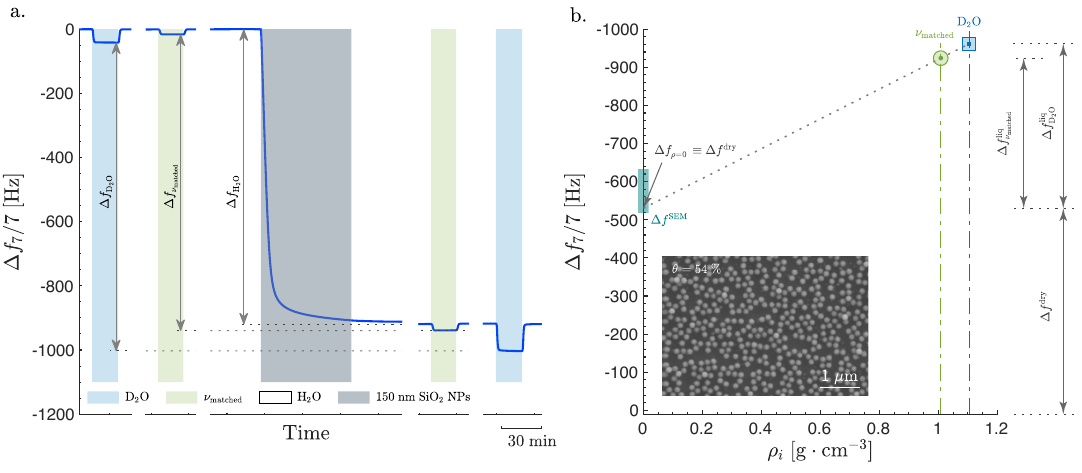}
	\caption {\textbf{Experimental procedure for measuring the QCM frequency response in different solutions and disentangling the adsorbate mass and the coupled liquid mass.} (a) The change in resonance frequency at the $7^{th}$ overtone versus time for: a flat poly-L-lysine (PLL)-coated substrate in contact with \ce{H2O}, \ce{D2O} and a glycerol/\ce{H2O} mixture matching the kinematic viscosity of \ce{D2O}, $\nu_{\mathrm{matched}}$; adsorption of $150 \; \mathrm{nm}$ \ce{SiO2} nanoparticles (NPs) until the jamming limit (corresponding to a surface coverage of $54\%$) is reached in an \ce{H2O}-based buffer; exchanging the solutions again to $\nu_{\mathrm{matched}}$ and \ce{D2O}. All solutions contained $10 \; \mathrm{mM}$ Bis-Tris, $\mathrm{pH \; (pD)} = 7.0$ and $150 \; \mathrm{mM}$ NaCl. A detailed experimental and calculation procedure is available in Sections S2-S4 and Figs.\ S1-S4 in the SI. (b) Frequency responses in \ce{D2O} and $\nu_\mathrm{matched}$ versus liquid density and a linear extrapolation towards vanishing density. $\Delta f^{\mathrm{SEM}}$ represents the calculated frequency response based on the surface coverage determined using SEM; using a NPs size of $143 \; \mathrm{nm}$ and density of $2.0 \pm 0.2 \; \mathrm{g \cdot cm^{-3}}$. Inset shows an SEM image of the surface of the sensor after drying; scale bar  $= 1 \; \mathrm{\mu m }$. A detailed imaging and counting procedure for SEM is available in Section S5 in the SI.}
	\label{fig:df_TOC}
	\end{center}
\end{figure}

Starting with high \ce{SiO2} NP surface coverage ($\theta = 54\%$), Fig.\ \ref{fig:df_TOC}a shows the three-step kinematic-viscosity matching procedure in which \ce{SiO2} NPs were electrostatically adsorbed to a smooth  poly-L-lysine (PLL)-coated \ce{SiO2} sensor. After initial establishment of a baseline in a buffered \ce{H2O}, the liquid in contact with the surface of the sensor was first replaced with buffered \ce{D2O} followed by buffered glycerol/\ce{H2O} mixture (4.55\%wt.\ glycerol) prepared to match the kinematic viscosity, $\nu =\eta/\rho$,  of \ce{D2O}, henceforth referred to as $\nu_{\mathrm{matched}}$. This was followed by adsorption of \ce{SiO2} NPs in an \ce{H2O}-based buffer until the $\Delta f$ response plateaued, after which the $\nu_{\mathrm{matched}}$ and \ce{D2O} buffers were again subsequently injected into the flow cell. As indicated in Fig.\ \ref{fig:df_TOC}a, this procedure enables measurements of the response induced upon NP adsorption in different liquids, $\Delta f_i$. 

A linear extrapolation of the frequency response in \ce{D2O}, $\Delta f_{\mathrm{D_2O}}$, and the kinematic-viscosity matched glycerol/\ce{H2O} mixture, $\Delta f_{\nu_{\mathrm{matched}}}$, plotted versus $\rho_i$ (Fig.\ \ref{fig:df_TOC}b) yields an intercept at $\rho=0$ of $\Delta f_{\mathrm{\rho=0}}  \approx -530 \; \mathrm{Hz} \; (\equiv 9540 \; \mathrm{ng \cdot cm^{-2}})$, a value which according to Eq.\ (\ref{eq:fmeasured}) represents the frequency shift of the dry mass of the adsorbed NPs, $\Delta f^{\rm dry}$. The obtained value is in excellent agreement with the independently determined dry mass of $10350 \pm 1035 \; \mathrm{ng \cdot cm^{-2}}$, based on \textit{ex-situ} SEM images obtained after drying the sensor under nitrogen flow (Fig.\ \ref{fig:df_TOC}b, inset). Note that the liquid contribution to $\Delta f_{\mathrm{H_2O}}$, representing the error introduced if  $\Delta f$ alone is used to estimate the dry mass of the adsorbate, is appreciable ($\approx -390 \; \mathrm{Hz}$ equivalent to $> 40\%$ of the measured response). Further, this separation of the contributions to the measured response shows that both dry mass of the adsorbate and the mass corresponding to dynamically coupled liquid can be quantified via self-sufficient measurements using mechanical resonators alone, thus resolving a severe limitation of the technique that has for long disqualified its use for quantitative mass-uptake analysis in liquid environments. Although in most practical situations, adsorption measurements are carried out as in this experiment, i.e., on a flat QCM crystal already immersed in a liquid, we have, additionally, verified the applicability of the method for measurements conducted with reference to air as well as for flat surfaces without adsorbates. For flat surfaces without adsorbates, the extrapolation to $\rho=0$ should yield $\Delta f_{\rho = 0} = 0$. Indeed the results presented in Section S6 in the SI shows that measurements conducted with reference to air yield the same dry mass as for measurements conducted with the QCM crystal already immersed in a liquid; it also shows that for surfaces without adsorbates the extrapolated results to $\rho=0$ yields $\Delta f_{\rho = 0} \approx 0$ (at all overtones), offering additional experimental verification to the kinematic viscosity matching approach and also providing an internal verification for the accuracy of the preparation of the kinematic viscosity-matched solutions by rendering an intercept at $\Delta f_{\rho = 0}=0$.

\begin{figure}[H] 
	\begin{center}
	\includegraphics[scale=1.0]{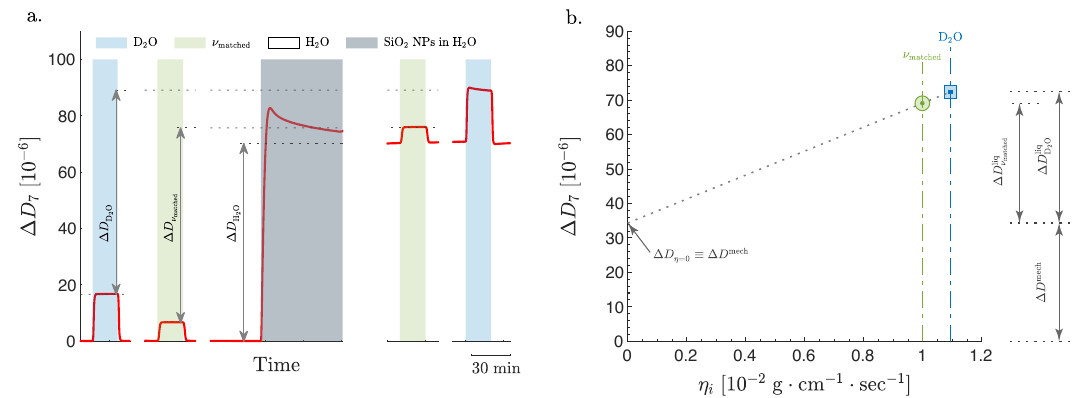}
	\caption {\textbf{Experimental procedure for measuring the QCM dissipation response in different solutions and disentangling the mechanical and liquid contributions.} (a) The change in dissipation at the $7^{th}$ overtone versus time for: a flat poly-L-lysine (PLL)-coated substrate in contact with \ce{H2O}, \ce{D2O} and glycerol/\ce{H2O} mixtures matching the kinematic viscosity of \ce{D2O}, $\nu_{\mathrm{matched}}$; the adsorption of $150 \; \mathrm{nm}$ \ce{SiO2} nanoparticles (NPs) until the jamming limit (corresponding to a surface coverage of $54\%$) is reached in an \ce{H2O}-based buffer; exchanging the solutions again to $\nu_{\mathrm{matched}}$ and \ce{D2O}. All solutions contained $10 \; \mathrm{mM}$ Bis-Tris, $\mathrm{pH \; (pD)} = 7.0$ and $150 \; \mathrm{mM}$ NaCl. A detailed experimental and calculation procedure is available in Sections S2-S4 in the SI. (b) Dissipation responses in \ce{D2O} and $\nu_{matched}$ versus liquid viscosity and a linear extrapolation towards vanishing viscosity.}
	\label{fig:dD_TOC}
	\end{center}
\end{figure}

\noindent
In the analysis above, the adsorbed NPs are assumed to be (i) rigid and (ii) firmly linked to the surface, i.e., the measured, $\Delta D_i$, should be solely due to the liquid contribution, $\Delta D^{\rm liq}_i$, with no mechanically dissipated energy, $\Delta D^{\rm mech} = 0$. For the \ce{SiO2} NPs under consideration, assumption (i) is valid. Concerning assumption (ii), the corresponding contribution to the dissipation, $\Delta D^\mathrm{mech}$, should be zero in order for the proposed approach to be strictly valid [see Eq.\ (\ref{eq:Dmeasured})]. It is therefore instructive to inspect the corresponding changes in $\Delta D$ (Fig.\ \ref{fig:dD_TOC}a) for the same sequence of injection events shown in Fig.\ \ref{fig:df_TOC}a.

Extrapolation of $\Delta D_\mathrm{D_2O}$ and the kinematic viscosity matched glycerol/\ce{H2O} mixture, $\Delta D_{\nu_\mathrm{matched}}$, to $\eta = 0$ has an intercept $\Delta D_{\mathrm{\eta=0}} \approx 34 \times 10^{-6}$ (Fig.\ \ref{fig:dD_TOC}b) which is equivalent to $-\Delta D_{\mathrm{\eta=0}}/\Delta f_{\mathrm{\rho=0}} = 0.06 \; \mathrm{[Hz^{-1}]}$ or $10^{-14}$ per NP and corresponds to $\approx 50\%$ of the measured $\Delta D_\mathrm{H_2O}$. The non-zero positive $\Delta D_{\mathrm{\eta=0}}$ was further confirmed by measurement conducted with reference to air as well as measurements of flat surfaces without adsorbate, the latter yielding $\Delta D_{\mathrm{\eta=0}} \approx 0$ (Section S6 in the SI). While $\Delta D_{\eta=0}=0$ for the NPs would have been an unequivocal evidence of rigidly adsorbed NPs, i.e., negligible viscoelastic contribution to the QCM response, this result shows that there is indeed a measurable mechanical contribution to the energy dissipation from the adsorbed NPs. For adsorbate systems with measurable viscoelastic character, $\Delta D_\mathrm{\eta=0}$ does not represent the dissipated energy under dry conditions, as in the case of perfectly rigid systems, but rather represents the mechanically dissipated energy in the adsorbate system surrounded by a liquid after excluding the energy dissipated in the liquid itself from the total dissipation response. The energy dissipated in the liquid is due to the corresponding viscous friction, which is referred to as $\Delta D^\mathrm{liq}$ in this work. As a matter of fact, the dissipated energy under experimentally dry conditions, i.e. air, is in almost all cases negligible (Fig.\ S5).  $\Delta D_\mathrm{\eta=0}$ in the case of rigid NPs is, thus, likely to reflect the mechanical energy dissipated due to the rotation and slippage of the NPs \cite{Ellis2004, Johannsmann2009}. Yet, the good agreement between the independent mass determination using SEM and the dry mass obtained by extrapolating $\Delta f_i$ in solutions matched with respect to kinematic viscosity (Fig.\ \ref{fig:df_TOC}b), suggests that the extrapolation approach is still applicable in spite of the positive $\Delta D_{\eta=0}$. In other words, the mechanical properties of the attachment between the NPs and the surface, reflected in a non-zero $\Delta D_{\eta=0}$, appears to have a negligible effect on $\Delta f_{\rho=0}$, thus suggesting that the kinematic-viscosity matching approach may be applicable also beyond perfectly rigid systems. We, therefore, aimed to investigate whether it is possible, using the measured response only, to directly validate if the mechanical properties of the contact between a specific adsorbate and the surface have a significant influence on the dry mass determination or not.

\begin{figure}[H] 
	\begin{center}
	\includegraphics[scale=1.0]{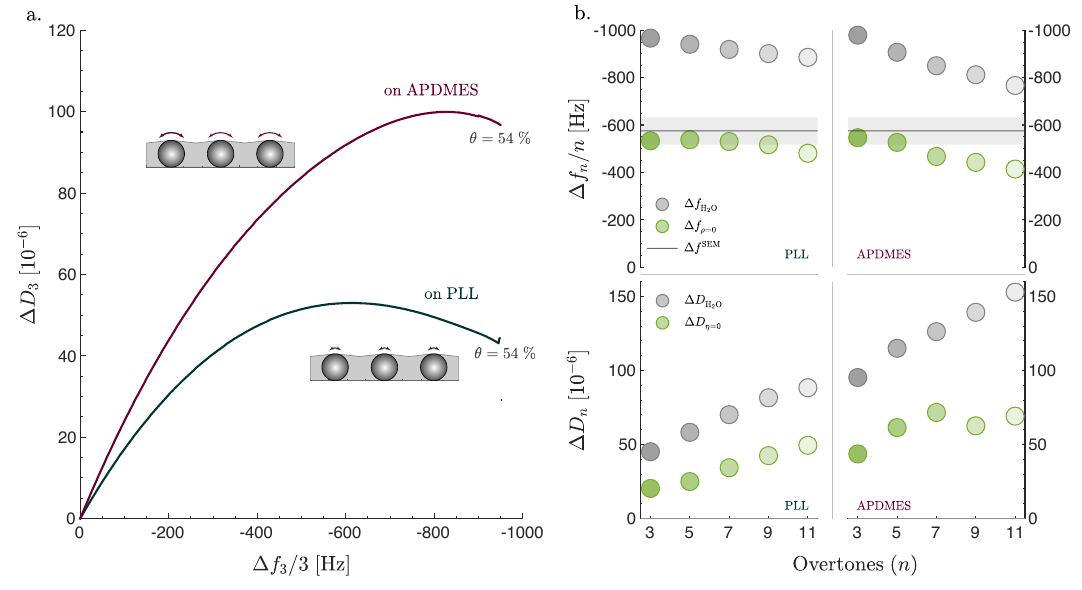}
	\caption {\textbf{Effect of the flexibility of the attachment between the NPs and the substrate on dry mass determination.} (a) Measured dissipation versus measured frequency at the $3^{rd}$ overtone for adsorption of $150 \; \mathrm{nm}$ \ce{SiO2} NPs on poly-L-lysine (PLL) -coated and on aminosilane (APDMES) -modified sensors. Two inset schematics illustrate a stronger rocking motion (represented by longer arrow on the top of each NP) for the NPs adsorbed to APDMES- than PLL-modified sensors. (b) Overtone dependence for the two experiments exhibited in panel (a), showing the measured frequency and dissipation responses in \ce{H2O} ($\Delta f_{\mathrm{H_2O}}$ and $\Delta D_{\mathrm{H_2O}}$) and the extrapolated frequency and dissipation responses ($\Delta f_{\mathrm{\rho = 0}}$ and $\Delta D_{\mathrm{\eta = 0}}$). $\Delta f^{\mathrm{SEM}}$ represents the calculated frequency response based on the surface coverage determined using SEM imaging; utilizing a NPs size of $143 \; \mathrm{nm}$ and density of $2.0 \pm 0.2 \; \mathrm{g \cdot cm^{-3}}$. The shaded area represents the uncertainty in the $\Delta f^{\mathrm{SEM}}$ which arises from the uncertainty in the density of the NPs. All solutions contained $10 \; \mathrm{mM}$ Bis-Tris, $\mathrm{pH \; (pD)} = 7.0$ and $150 \; \mathrm{mM}$ NaCl.}
	\label{fig:Linker_Stiffness}
	\end{center}
\end{figure}

To address this question, we performed experiments where the stiffness of the attachment between NPs and the underlying surface was deliberately varied by replacing the PLL-coated surfaces with a surface modified with aminosilane (APDMES). Binding of NPs to the surface modified with APDMES resulted in significantly higher $\Delta D/\Delta f$ ratio than the binding to PLL did (Fig.\ \ref{fig:Linker_Stiffness}a), being indicative of a weaker NP-surface interactions for the APDMES-modified surface (see Section S7 of the SI.) Furthermore, values of $\Delta f_{\mathrm{H_2O}}$ had a more pronounced overtone dependence for the APDMES-modified surface than for the PLL-coated surface (Fig.\ \ref{fig:Linker_Stiffness}b, grey markers). Notably, while the overtone dependence essentially disappeared for the extrapolated $\Delta f_{\mathrm{\rho=0}}$ for NPs bound to the PLL-coated surface, a clear overtone dependence was observed for NPs bound to the APDMES-modified surface (Fig.\ \ref{fig:Linker_Stiffness}b, green markers). This is expected since the adsorbate-induced $|\Delta f|$ typically decreases with increasing overtone number, $n$. This phenomenon arises from two different sources. The first is a viscoelastic contribution (refereed to as softness in some previous work), which results in decreasing $|\Delta f|$ with increasing $n$ \cite{Johannsmann2008b}. The second is from an $n$-dependent liquid response, i.e., $|\Delta f|$ also decreases with increasing $n$. Notably, the $n$-dependent liquid response will also occur for rigidly coupled adsorbates; this has been previously shown through theoretical models  \cite{Johannsmann2008b} as well as through experimental results of rough electrodeposited copper adlayers \cite{Friedt2003, Friedt2003a}. Hence, for systems with a negligible viscoelastic influence on $\Delta f$, the overtone dependence of $\Delta f$ will thus arise primarily from the liquid contribution and therefore the overtone dependence is expected to vanish if the liquid contribution is eliminated using the extrapolation approach. The observed difference in the overtone dependence between APDMES and PLL for $\Delta f_{\mathrm{\rho=0}}$ can thus be attributed to a higher viscoelastic contribution upon NP adsorption on APDMES-modified than upon PLL-coated surfaces. This conclusion is further supported by the higher $\Delta D_{\mathrm{H_2O}}$ and $\Delta D_{\mathrm{\eta=0}}$ values measured for the APDMES-coated surface where both the value of $\Delta D_{\mathrm{\eta=0}}$ and the deviation between $\Delta f_{\mathrm{\rho=0}}$ and the dry mass determined using SEM decrease with decreasing $n$ (Fig.\ \ref{fig:Linker_Stiffness}b), suggesting a weaker contribution from viscoelastic losses to $\Delta f_{\mathrm{\rho=0}}$ at lower $n$. In fact, the dry mass of NPs adsorbed to the APDMES-modified surface determined from $\Delta f_{\mathrm{\rho=0}}$ at $n=3$ agreed perfectly with the SEM-determined value, within the uncertainty limits of the SEM approach. The overtone dependence on the linker stiffness has been further verified using two sets of experiments: (i) carboxylated Au NPs on APDMES at different ionic strengths (Fig.\ S13) and (ii) PEGylated Au NPs with different PEG linker lengths on APDMES (Fig.\ S13). These experiments clearly demonstrate the increasing overtone dependence of the measured and extrapolated frequency response with decreasing linker stiffness. It also shows that for all the experiments, even for Au NPs with 5000 Da PEG linker, the extrapolated frequency response at $n=3$ accurately estimates the dry mass of the NPs within less than 10\% error compared to the complementary measurements of QCM in air.

\begin{figure}[H] 
	\begin{center}
	\includegraphics[scale=1.0]{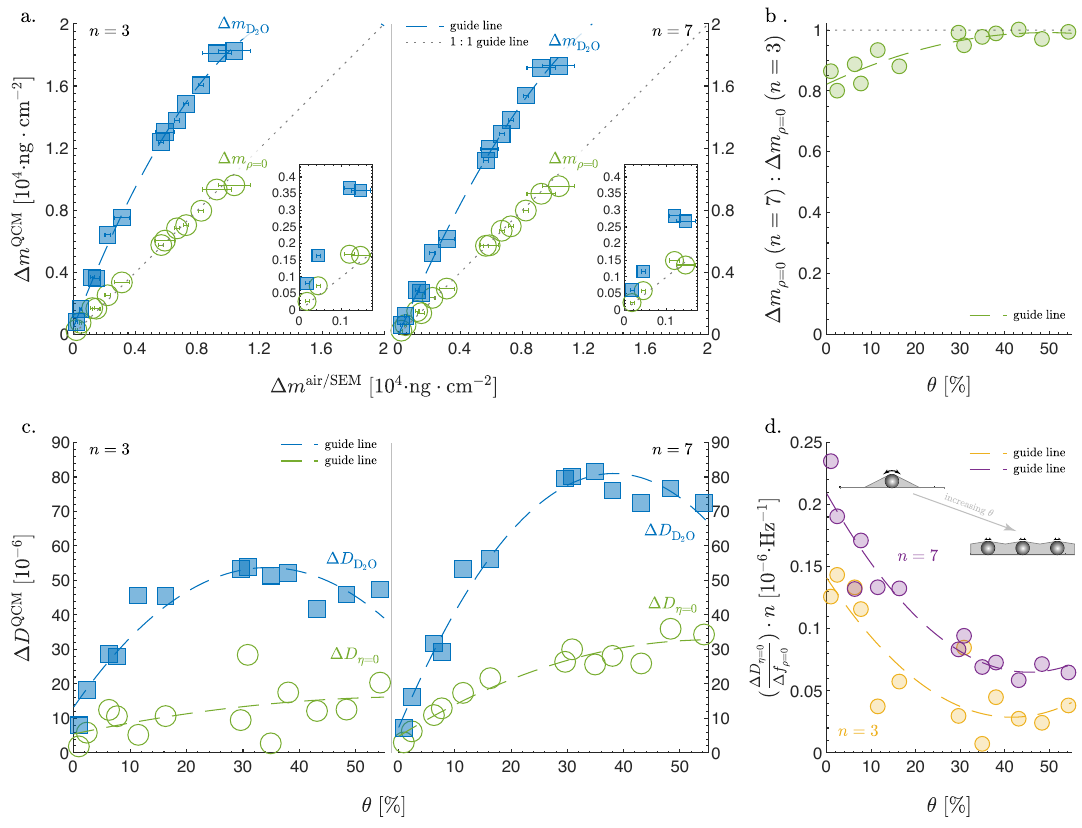}
	\caption {\textbf{Disentangling the liquid and adsorbate mass/mechanical contributions to the QCM response of adsorbed \ce{SiO2} nanoparticles over a broad range of surface coverages.} (a) Comparison between the measured QCM mass in \ce{D2O}, $\Delta m_{\mathrm{D_2O}}$, and the mass obtained by extrapolating the responses in \ce{D2O} and in glycerol/\ce{H2O} mixture matching the kinematic viscosity of \ce{D2O}, $\nu_{\rm matched}$, to $\rho = 0$, $\Delta m_{\mathrm{\rho =0}}$, versus the mass determined using SEM imaging or frequency response in air. Error bars represent the uncertainty in the air or SEM measurements as detailed in Section S5 in the SI. The QCM mass was calculated using the Sauerbrey equation for $5 \; \mathrm{MHz}$ sensors: $\Delta m = -18.0 \cdot \Delta f$. Results are shown for the $3^{rd}$ and $7^{th}$ overtone ($n=3$ and $7$); results from all overtones are presented in Fig.\ S16. Insets show a zoomed-in portion of the plot in the low adsorbed mass region. (b) Ratio of the mass determined using the results of the $7^{th}$ overtone to that of the $3^{rd}$ overtone versus surface coverage. (c) Comparison between the measured QCM energy dissipation in \ce{D2O}, $\Delta D_{\mathrm{D_2O}}$, and the mechanically dissipated energy obtained by extrapolating the responses in \ce{D2O} and in $\nu_{\rm matched}$ to $\eta = 0$, $\Delta D_{\mathrm{\eta =0}}$, versus the surface coverage. Results are shown for the $3^{rd}$ and $7^{th}$ overtone; results from all overtones are given in Fig.\ S17. (d) Mechanical energy dissipated normalized by the frequency response of the dry mass for the NPs versus surface coverage at the $3^{rd}$ and $7^{th}$ overtone. All solutions contained $10 \; \mathrm{mM}$ Bis-Tris, $\mathrm {pH \; (pD)} = 7$, and $150 \; \mathrm{mM}$ \ce{NaCl}.}
	\label{fig:theta}
	\end{center}
\end{figure}

An additional long-standing question is to what extent QCM data reflect lateral interactions between closely-packed adsorbates on a surface. In particular, as seen in Figs.\ \ref{fig:dD_TOC}a and \ref{fig:Linker_Stiffness}a, $\Delta D$ reaches a transient maximum (peak) and starts decreasing again as more NPs adsorb. This transient peak has been previously ascribed to hydrodynamic stabilization of NPs, i.e., less rotation and slippage, that becomes appreciable when NPs are sufficiently close to each other  \cite{Johannsmann2009}, i.e., at high-enough surface coverage. We were thus curious to inspect whether the kinematic-viscosity matching approach applied for a varying surface coverage ($\theta$) of NPs could contribute new insight into this issue. Both the dry mass of NPs, $\Delta m_{\mathrm{\rho=0}}$, and the mechanically dissipated energy associated with the contact zone between NPs and the surface, $\Delta D_{\mathrm{\eta=0}}$, were therefore determined at various surface coverages, $\theta$, ranging from $1$ to $54 \%$, (Fig.\ \ref{fig:theta}a,c). The mass determined using the extrapolation approach, $\Delta m_{\mathrm{\rho=0}}$ calculated based on $\Delta f_{\mathrm{\rho=0}}$ using Eq.\ (\ref{eq:Sauerbrey}), and the mass determined using QCM in air or SEM, $\Delta m^{\mathrm{air/SEM}}$, are indeed in almost perfect agreement (Fig.\ \ref{fig:theta}a) over the entire coverage regime. The overtone dependence of $\Delta m_{\mathrm{\rho=0}}$ was fairly weak: the ratio between the seventh to the third overtone values of $\Delta m_{\mathrm{\rho=0}}$ ranged from $\approx 80\%$ at the lowest coverage to higher than $95\%$ above $30\%$ coverage (Fig.\ \ref{fig:theta}a-b), demonstrating that the kinematic viscosity approach can be used to successfully disentangle the dry and liquid mass in a broad coverage regime. Further, the transient peak in the measured energy dissipation, $\Delta D_{\mathrm{D_2 O}}$, disappears in the corresponding curve for the extrapolated $\Delta D_{\mathrm{\eta=0}}$ (Fig.\ \ref{fig:theta}c), revealing that the changes in liquid-related energy dissipation versus surface coverage is the source of the observed transient peak in the measured dissipation. We also observed that $\Delta D_{\mathrm{\eta=0}}$ contributes only between $30$ and $50 \%$ to the measured dissipation at all surface coverages, i.e., the larger part of the dissipated energy is liquid-related in nature. Additionally, the contribution from the mechanically dissipated energy normalized by the extrapolated frequency response, $\Delta D_{\mathrm{\eta=0}}/\Delta f_{\mathrm{\rho=0}}$, decreases with increasing $\theta$ (Fig.\ \ref{fig:theta}d), demonstrating that the mechanical energy dissipated per NP decreases with increasing $\theta$. These observations show that the NPs are indeed mechanically stabilized with increasing $\theta$, in good agreement with previous finite element modelling (FEM) of similar systems \cite{Johannsmann2009}; a conclusion also confirmed by the observed decrease in overtone dependence of $\Delta f_{\mathrm{\rho=0}}$ with increasing $\theta$ (Fig.\ \ref{fig:theta}b). This analysis thus shows that the kinematic-viscosity matching approach can be used not only to accurately determine the dry mass of the adsorbate over a broad coverage range, but also to experimentally quantify the mechanically dissipated energy, opening up an new avenue to access various material and/or linker properties.

\begin{figure}[H] 
	\begin{center}
	\includegraphics[scale=1.0]{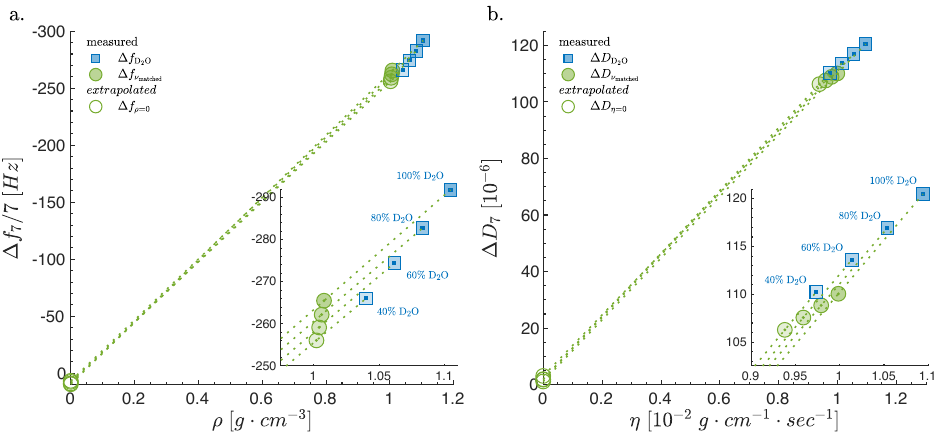}
	\caption {\textbf{Extrapolation from \ce{D2O}/\ce{H2O} mixtures and their kinematic viscosity matched glycerol/\ce{H2O} mixtures.} (a) Frequency versus liquid density and (b) dissipation versus liquid viscosity of four different \ce{D2O}/\ce{H2O} mixtures and their kinematic viscosity matched glycerol/\ce{H2O} mixtures. The responses represent exchanging the medium in contact with a flat substrate from air to the different liquids. Shown as well are the extrapolation lines to $\rho =0$ and $\eta = 0$. $100\%$, $80\%$, $60\%$, and $40\%$ \ce{D2O} correspond to the molar fractions $1.0$, $0.8$, $0.6$, and $0.4$ of \ce{D2O}. Results were obtained from measurements on an aminosilane-modified sensor using citrate-based buffer, $\mathrm{pH \; (pD)} = 4.0$.}
	\label{fig:D2OH2OMix}
	\end{center}
\end{figure}

So far we have relied on complementary methods, i.e., QCM in air and SEM, to verify the accuracy of the extrapolation approach. The objective of this work is to obtain accurate, trustworthy results while obviating the need for any complementary methods. As for any other scientific method, obtaining reliable data can only be achieved by conducting replicas and assessing the variability from one measurement to another. However, in some cases it is very challenging to reproduce the same exact experimental conditions and thus outcome, such as for example the same exact surface coverage of an adsorbate. By using more than one \ce{D2O}/\ce{H2O} mixture and their matched \ce{H2O}/glycerol mixture to estimate the dry mass and mechanical properties, one can extend the kinematic viscosity matching approach to provide additional confidence to the results obtained from a single measurement. Fig.\ \ref{fig:D2OH2OMix} shows the responses measured upon exchanging the medium in contact with a flat substrate from air to four different \ce{D2O}/\ce{H2O} mixtures and their corresponding kinematic viscosity matched glycerol/\ce{H2O} mixtures. Since there is no adsorption taking place in this process, the dry mass and the mechanical dissipated energy should be zero. Indeed Fig.\ \ref{fig:D2OH2OMix} shows that the extrapolation from the four different pairs of kinematic viscotsity matched solutions yields $\Delta f_\mathrm{\rho=0} \approx 0$ and $\Delta D_\mathrm{\eta=0} \approx 0$. We have applied this approach on flat substrates without adsorbate (Figs.\  \ref{fig:D2OH2OMix},S18), 150 nm \ce{SiO2} NPs at different surface coverages (Fig.\ S19), 70 nm Au NPs at different ionic strengths (Fig.\ S20), and 70 nm PEGylated Au NPs with different PEG linker lengths (Fig.\ S21). The results acquired from all these experiments show that $\Delta f_\mathrm{\rho=0}$ and $\Delta D_\mathrm{\eta=0}$, obtained by extrapolation of the response in four different \ce{D2O}/\ce{H2O} mixtures with their corresponding kinematic viscosity matched glycerol/\ce{H2O} mixtures to $\rho=0$ and $\eta=0$, all converge to similar values. The results showed systematic dependence for neither $\Delta f_{\mathrm{\rho=0}}$ nor $\Delta D_{\mathrm{\eta=0}}$ on the \ce{D2O} to \ce{H2O} mixing ratios. On one side, these results provide further verification for the kinematic viscosity matching approach to determine the adsorbate mass and the mechanical properties of the adsorbate/surface interactions. This additional evidence verifies indeed the accuracy of $\Delta f_\mathrm{\rho=0}$ to determine the adsorbate mass, in agreement with the evidence provided provided based on the results from QCM in air and SEM imaging. It is, however, a more crucial control in the case of $\Delta D_\mathrm{\eta=0}$, because there is limited existing knowledge with respect to the nature of the mechanically dissipated energy and whether it is affected by the properties of the liquid. These results suggest that the mechanically dissipated energy was not affected by the different liquid properties within the range of liquids used in this work. If there were a dependency on the liquid properties one would have expected a systematic increase or decrease in $\Delta D_\mathrm{\eta=0}$ with the increasing \ce{D2O} concentrations in the \ce{D2O}/\ce{H2O} mixtures. On the other side, the possibility to determine $\Delta f_\mathrm{\rho=0}$ and $\Delta D_\mathrm{\eta=0}$ based on the extrapolation of four couples of data points provides a robust statistical verification of the analysis especially for systems with small $\Delta f$ and $\Delta D$ responses. Note that in this work we determine the dry mass and mechanically dissipated energy at $\theta = 1\%$ of 150 nm \ce{SiO2} NPs, which is equivalent to $\Delta f_\mathrm{\rho=0} \approx -15 \; \mathrm{Hz}$ and $\Delta D_\mathrm{\eta=0} \approx 2 \cdot 10^{-6}$; $-15 \; \mathrm{Hz}$ corresponds to the mass of monolayer of small protein of $\approx 5 \; \mathrm{nm}$ in diameter at the jamming limit ($\theta = 54\%$). Finally, it is also worth noting that one could achieve a sound statistical verification of $\Delta f_\mathrm{\rho=0}$ and $\Delta D_\mathrm{\eta=0}$, as the one achieved using different \ce{D2O}/\ce{H2O} mixtures, by exchanging the same kinematic viscosity matched liquids, e.g., \ce{D2O} and 4.55\%wt glycerol in \ce{H2O}, several times before and after the adsorbate.

\begin{figure}[H] 
	\begin{center}
	\includegraphics[scale=1.0]{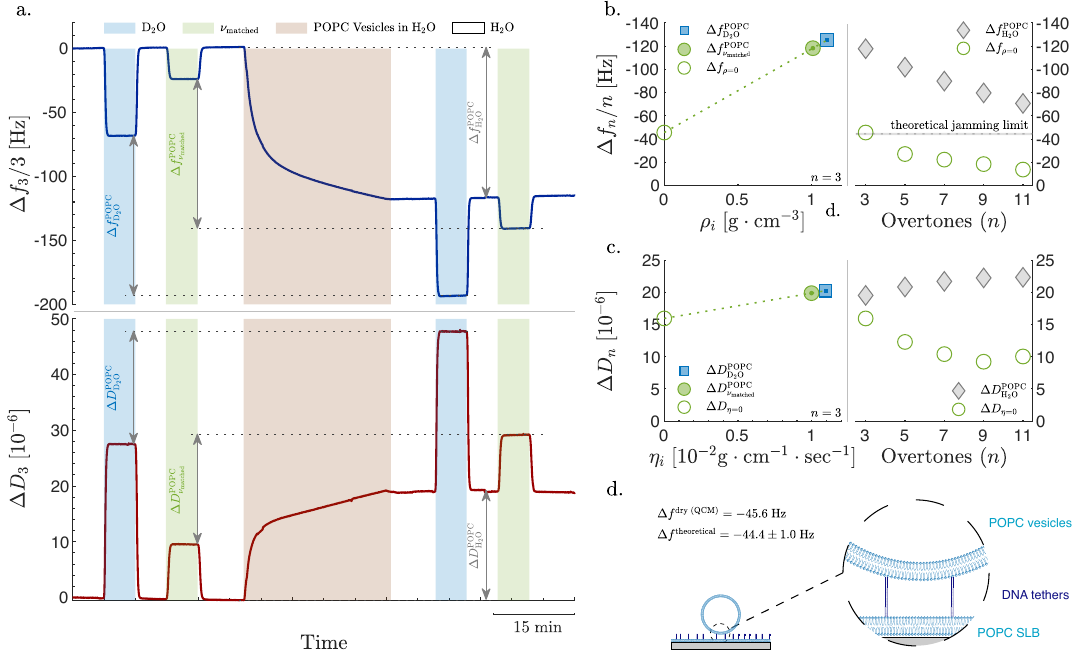}
	\caption {\textbf{Disentangling the liquid and adsorbate mass/mechanical contributions to the QCM response of a tethered lipid vesicles.}  (a) The change in resonance frequency (upper panel) and dissipated energy (lower panel) at the $3^{rd}$ overtone versus time for the tethering of POPC lipid vesicles (size distribution shown in Fig.\ S23) to a supported lipid bilayer (SLB) through DNA molecules with cholesterol anchors. The buffer was exchanged to \ce{D2O}-based buffer and glycerol/\ce{H2O} mixture matching the kinematic viscosity of \ce{D2O}, $\nu_{\mathrm{matched}}$, before and after the POPC adsorption step. A complete experimental procedure including the formation of SLB and attachment of the DNA tethers is shown in Fig.\ S23. (b) Frequency responses (left panel) for the POPC vesicles in \ce{D2O} and $\nu_{matched}$ versus liquid density and a linear extrapolation towards vanishing density, $\rho =0$. Overtone dependence (right panel) for the measured frequency responses in \ce{H2O}, $\Delta f_{\mathrm{H_2O}}$, and the extrapolated frequency responses, $\Delta f_{\mathrm{\rho=0}}$, of the POPC vesicles.  The theoretical frequency response, $\Delta f^{\mathrm{theoretical}}$, was calculated using the Sauerbrey equation for $5 \; \mathrm{MHz}$ sensors: $\Delta m = -18.0 \cdot \Delta f$ based on POPC molecular weight of $760.076 \; \mathrm{g\cdot mol^{-1}}$, area per lipid of $0.683 \pm 0.015 \; \mathrm{nm^2}$ (Ref. \citenum{Kucerka2006}), and assumed surface coverage of $\theta = 54\%$.  (c) Dissipation responses (left panel) for the POPC vesicles from \ce{D2O} and $\nu_{matched}$ versus liquid viscosity and a linear extrapolation towards vanishing viscosity, $\eta =0$. Overtone dependence (right panel) for the measured dissipation responses in \ce{H2O}, $\Delta D_{\mathrm{H_2O}}$, and the extrapolated dissipation, $\Delta D_{\mathrm{\eta=0}}$, of the POPC vesicles. (d) A schematic showing POPC vesilces tethered to SLB via DNA molecules. Experimental details, characterization of vesicles and calculations are presented in Section S11 in the SI. All solutions contained $10 \; \mathrm{mM}$ phosphate buffer, $\mathrm{pH \; (pD)} = 7.4$, $2.7 \; \mathrm{mM}$ \ce{KCl}, and $137 \; \mathrm{mM}$ \ce{NaCl}.}
	\label{fig:POPC}
	\end{center}
\end{figure}

Encouraged by the positive results obtained using the kinematic viscosity matching approach applied on rigid NPs even under weak attachment conditions, we decided to explore the applicability of the approach to less rigid, biological NPs. As a proof of concept we determine the dry mass of POPC vesicles anchored to a supported lipid bilayer (SLB) via DNA tethers (Fig.\ \ref{fig:POPC}a,d). As previously described \cite{Benkoskr2005}, DNA tethers are formed of 45 base pairs with a stretched length of $\approx 15 \; \mathrm{nm}$ and have two cholestrol anchors at its ends enabling self-insertion of one end into the SLB and the other into the POPC vesicles. The flexibility of the DNA linkers and the POPC vesicles suggests that the system will have an appreciable viscoelastic character; indeed the measured $\Delta f_\mathrm{H_2O}$ showed a very strong overtone dependence ranging from $-120 \; \mathrm{Hz}$ at $n = 3$ to $-70 \; \mathrm{Hz}$ at $n = 11$ (Fig.\ \ref{fig:POPC}b) and the $-\Delta D_\mathrm{H_2O}/\Delta f_\mathrm{H_2O}$ ratio ranged from 0.17 at $n = 3$ to 0.32 for $n = 11$ (Fig.\ \ref{fig:POPC}b-c). Extrapolation of the frequency response in \ce{D2O} and $\nu_{\mathrm{matched}}$ buffers towards $\rho = 0$, yields an intercept $\Delta f_{\mathrm{\rho=0}}$ of $-45.6 \; \mathrm{Hz}$ at $n=3$ with strong overtone dependence (Fig.\ \ref{fig:POPC}b). $\Delta f_{\mathrm{\rho=0}}$ at $n=3$ is in excellent agreement with the theoretically estimated frequency response. The theoretical frequency response was estimated using the Sauerbrey equation [Eq. (\ref{eq:Sauerbrey})] based on POPC molecular weight of $760.076 \; \mathrm{g\cdot mol^{-1}}$, area per lipid of $0.683 \pm 0.015 \; \mathrm{nm^2}$ (Ref. \citenum{Kucerka2006}), and surface coverage of $\theta = 54\%$. The assumed surface coverage is justified by the plateauing adsorption profile of the POPC vesicles (Fig.\ \ref{fig:POPC}a), which indicates that the surface coverage is very close to the jamming limit. Notably the theoretical frequency response is independent of the size of the vesicles as detailed in Section S11 in the SI. It is worth mentioning that $\Delta f_{\mathrm{\rho=0}}$ determined here represents the dry mass of the lipids only in the POPC vesicles without the mass of the liquid in their core. This is due to the fact that both \ce{D2O} and glycerol are permeable to POPC lipid vesicles; thus the liquid inside the vesicles are also exchanged when exchanging the bulk liquid. From a theoretical perspective, the liquid inside the vesicles is in this case is included in the Navier-Stokes-related liquid contributions, i.e., $k_f \cdot \rho_i$ in Eq.\ (\ref{eq:fmeasured}). If one instead would use two liquids that are impermeable to the vesicles, then $k_f \cdot \rho_i$ would not include the liquid inside the vesicles and thus $\Delta f_{\mathrm{\rho=0}}$ would represent the mass of the lipids and the liquid inside the vesicles. Such distinction is very important for correct data interpretation. Finally, by inspecting $\Delta D_\mathrm{\eta = 0}$ at $n=3$, we observe that most of the energy dissipated is mechanical in nature, which is in agreement with the expected strong viscoelastic character of the adsorbate system. Note that $\Delta D_\mathrm{\eta = 0}$ does not only reflect the mechanical properties of the linker, as in the cases of \ce{SiO2} and Au NPs, but also reflects the mechanical properties of the vesicles themselves. Altogether, these results show that the kinematic viscosity matching approach can provide valuable information beyond rigid systems to systems with appreciable viscoelastic character.

\begin{figure}[H] 
	\begin{center}
	\includegraphics[scale=1.0]{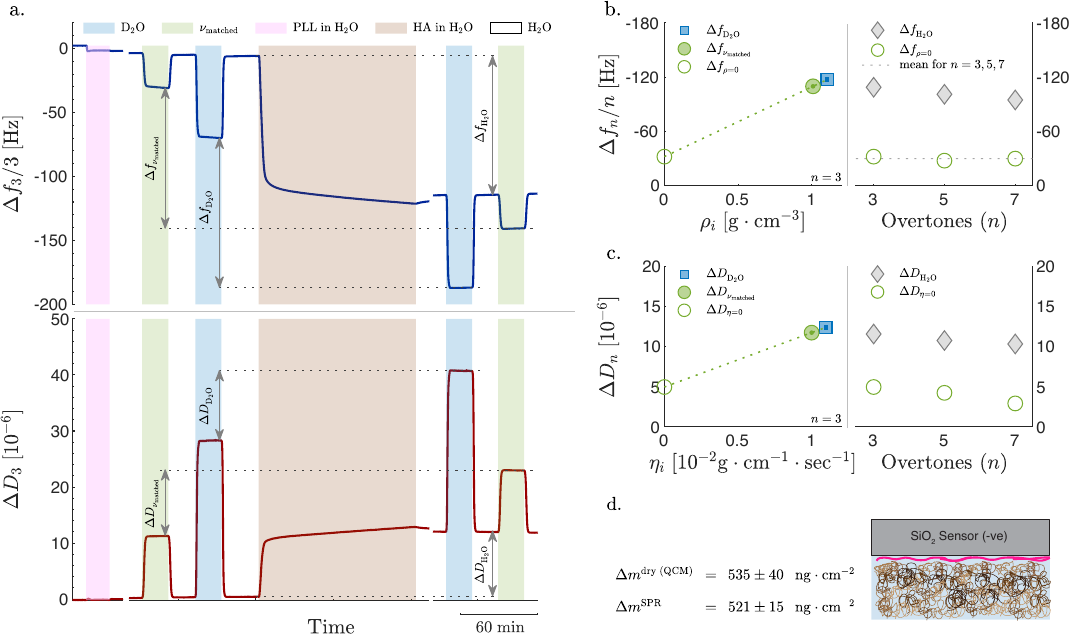}
	\caption {\textbf{Disentangling the liquid and adsorbate mass/mechanical contributions to the QCM response of a polymer adlayer.}  (a) The change in resonance frequency (upper panel) and dissipated energy (lower panel) at the $3^{rd}$ overtone versus time for the adsorption of poly-L-lysine (PLL) adlayer on bare \ce{SiO2} sensor, followed by deposition of a humic acids (HA) adlayer. The buffer was exchanged to \ce{D2O}-based buffer and glycerol/\ce{H2O} mixture matching the kinematic viscosity of \ce{D2O}, $\nu_{\mathrm{matched}}$, before and after the HA adsorption step. (b) Frequency responses (left panel) for the HA layers from \ce{D2O} and $\nu_{matched}$ versus liquid density and a linear extrapolation towards vanishing density, $\rho =0$. Overtone dependence (right panel) for the measured frequency responses in \ce{H2O}, $\Delta f_{\mathrm{H_2O}}$, and the extrapolated frequency responses, $\Delta f_{\mathrm{\rho=0}}$, of the HA adlayer. Results are shown for $n=3,5,$ and $7$; results from higher overtone numbers could not be utilized due to unstable instrumental signal. (c) Dissipation responses (left panel) for the HA layers from \ce{D2O} and $\nu_{matched}$ versus liquid viscosity and a linear extrapolation towards vanishing viscosity, $\eta =0$. Overtone dependence (right panel) for the measured dissipation responses in \ce{H2O}, $\Delta D_{\mathrm{H_2O}}$, and the extrapolated dissipation, $\Delta D_{\mathrm{\eta=0}}$, of the HA adlayer. Results are shown for $n=3,5,$ and $7$; results from higher overtone numbers could not be utilized due to unstable instrumental signal. (d) A schematic showing the HA adlayer. The dry mass, $\Delta m^{\mathrm{dry \;(QCM)}}$, was calculated using the Sauerbrey equation for $5 \; \mathrm{MHz}$ sensors: $\Delta m = -18.0 \cdot \Delta f$ based on $\Delta f_{\mathrm{\rho=0}}$ averaged from $n=3,5,$ and $7$. $\Delta m^\mathrm{SPR}$ was determined as detailed in Section S12 in the SI. All solutions contained $10 \; \mathrm{mM}$ Bis-Tris, $\mathrm{pH \; (pD)} = 7.0$, $150 \; \mathrm{mM}$ \ce{NaCl} and $2.5 \; \mathrm{mM}$ \ce{CaCl2}.}
	\label{fig:HA}
	\end{center}
\end{figure}

The principles introduced above to determine the dry mass of adsorbed NPs are expected to also apply to adsorbed polymer films  \cite{Daikhin1996}, given that they are in a regime with sufficiently small viscoelastic contributions, i.e., where the Sauerbrey equation [Eq.\ (\ref{eq:Sauerbrey})] is applicable not only for the dry mass but also the liquid coupled to the adsorbed polymer. This was here tested by investigating electrostatically-driven adsorption of HA to a PLL-coated sensor (Fig.\ \ref{fig:HA}a), using the aforementioned strategy of kinematic viscosity matching. Extrapolation of the frequency response in \ce{D2O} and $\nu_{\mathrm{matched}}$ buffers towards $\rho = 0$, yields an intercept $\Delta f_{\mathrm{\rho=0}}$ of $-29.8 \pm 2.2 \; \mathrm{Hz}$ with essentially no overtone dependence for $\Delta f_{\mathrm{\rho=0}}$ values (Fig.\ \ref{fig:HA}b). Additionally, conversion of $\Delta f_{\mathrm{\rho=0}}$ into mass using Eq.\ (\ref{eq:Sauerbrey}) perfectly agrees with the mass obtained by employing complementary SPR measurements (see Fig.\ \ref{fig:HA}d and Section S12 in the SI), confirming the validity of the extrapolation approach to determine the dry mass. Additionally, the mechanically dissipated energy represented by $\Delta D_{\mathrm{\eta=0}}$ is less than half of the total dissipation, demonstrating that the majority of the dissipated energy is liquid-related in nature. For the applicability of the Sauerbrey equation on polymeric adlayers, it has been suggested that the dissipation-to-frequency ratio should be much smaller than $0.4 \times 10^{−6} \; \mathrm{Hz^{−1}}$ [$\Delta D/\Delta f \ll 0.4 \times 10^{−6} \; \mathrm{Hz^{−1}}$  \cite{Reviakine2011}]. Although $\Delta D_{\mathrm{H_2O}} / \Delta f_{\mathrm{H_2O}}$, upon HA adsorption ($\approx 0.11 \times 10^{-6} \; \mathrm{Hz}^{-1}$) is indeed smaller than this threshold, it is on the same order, and thus does not completely meet this criterion. Still the perfect agreement of $\Delta f_{\mathrm{\rho=0}}$ with the mass determined using SPR, suggests that the kinematic viscosity matching approach is applicable beyond perfectly rigid systems also for polymeric adlayers as for NPs.

In conclusion, our results demonstrate that the proposed kinematic viscosity matching approach makes it possible to determine the molecular mass of both nanoscale adsorbates of arbitrary shape as well as polymeric adlayers, even for systems with appreciable viscoelastic contributions, and that the mechanically dissipated energy associated with the adsorbate and/or its attachment can be specifically quantified. In particular, the results show that the dry-mass determination can be self-sufficiently validated based on the overtone dependence of the extrapolated $\Delta f_{\mathrm{\rho=0}}$. For systems with weak to no measurable overtone dependence, the viscoelastic contribution is negligible and thus the true dry mass can be determined based on any of the measured overtones. For systems with an appreciable overtone-dependence, the mass determination at the lowest overtone number is the most reliable, albeit the full applicability for such systems merits further investigations. In addition, the approach does not require any \textit{a priori} knowledge about the adsorbed material, which makes it unique in comparison with most commonly applied \textit{in situ} methods, where ellipsometry and SPR stand out as the most reliable and commonly used methods. However, for accurate mass determination using these methods, at least the derivative of the refractive index with respect to the concentration of the adsorbate, $dn/dc$, must be known or independently determined. Additionally, in the case of SPR, also the film thickness and the decay length of the evanescent sensing field have to be determined independently. Finally, the possibility to simultaneously determine the dry mass and mechanical properties of nano-sized adsorbates as well as the liquid contributions using a single measurement and on the same spatial scale provides the unique opportunity to potentially extract more detailed information about the adsorbates, such as interrogating orientation, spatial distribution, and binding strength of adsorbates. Additionally, although here demonstrated using QCM, the applicability of the kinematic viscosity matching approach goes beyond QCM to a variety of mechanical resonators that operate in liquid environment, and can thus potentially be utilized for single cell \cite{Martinez-Martin2017}  and even single virus \cite{Dolai2019} mass determination.

\newpage

\section*{Materials and Methods}
\subsection*{Chemicals \& Materials}
Ethanol ($\geq 99.5 \%$) was purchased from Solveco (Roserberg, Sweden). Sodium dodecyl sulfate (SDS; $\geq 99.0 \%$), anhydrous methanol ($99.8 \%$), glycerol ($\geq 99 \; \%$), \ce{D2O} ($99.9 \; atom \; \% \; \mathrm{D}$), \ce{DCl} ($\geq 99 \; atom \; \% \; D$), \ce{NaOD} ($40 \; \% \mathrm{wt.}$), \ce{HCl} ($1 \; M$), \ce{NaCl} ($\geq 99 \%$), \ce{CaCl2} anhydrous ($\geq 93 \%$), phosphate buffered saline (PBS) tablets, bis(2-hydroxyethyl)amino-tris- (hydroxymethyl)-methane (Bis-Tris; $\geq 98 \%$), citric acid monohydrate ($\geq 98 \%$), trisodium citrate dihydrate, poly-L-lysine (PLL; $\mathrm{MW} = 70000$ -- $150000$), humic acids (HA), (3-aminopropyl)dimethylethoxysilane, $97\%$, $150 \; \mathrm{nm}$ \ce{SiO2} nanoparticles (NPs), anhydrous chloroform ($\geq99 \%$),  1-palmitoyl-2-oleoyl-glycero-3-phosphocholine (POPC), gold(III) chloride hydrate (\ce{HAuCl4 . xH2O}, 99.999\%), hydroquinone ($\ge 99\%$), O-(2-Carboxyethyl)-O'-(2-mercaptoethyl)heptaethylene glycol ($>95\%$; MW = 458.6 Da; 500 PEG), and 2100 Da [Poly(ethylene glycol) 2-mercaptoethyl ether acetic acid (2000 PEG), were purchased from Merck Sigma-Aldrich (Darmstadt, Germany). 5000 Da (alfa-carboxy-beta-mercapto-PEG; 5000 PEG) were purchased from RAPP Polymere GmbH (T\"ubingen, Germany). The following cholesteryl-TEG modified DNA oligonucleotides, modified with a  were purchased from Eurogentec (Seraing, Belgium).

\begin{itemize}
    \item 5'-TGG-ACA-TCA-GAA-ATA-AGG-CAC-GAC-GGA-CCC-3'-TEG-Cholesterol (ssA)
    \item Cholesterol-TEG-5'-CCC-TCC-GTC-GTG-CCT-3' (ssB)
    \item 5'-TAT-TTC-TGA-TGT-CCA-AGC-CAC-GAG-TTC-CCC-3'-TEG-Cholesterol (ssC)
    \item Cholesterol-TEG-5'-CCC-GAA-CTC-GTG-GCT-3' (ssD)
\end{itemize}{}

\subsection*{Buffers}
Three different buffers were used 
\begin{itemize}
    \item $\mathrm{pH \; (pD)} = 7.0$, $150 \; \mathrm{mM}$ \ce{NaCl}, and $10 \; \mathrm{mM}$ Bis-Tris
    \item $\mathrm{pH \; (pD)} = 7.0$, $150 \; \mathrm{mM}$ \ce{NaCl}, $2.5 \; \mathrm{mM}$ \ce{CaCl2}, and $10 \; \mathrm{mM}$ Bis-Tris
    \item $\mathrm{pH \; (pD)} = 4.0$, $\approx 3.8 \; \mathrm{mM}$ trisodium citrate dihydrate and $\approx 6.2 \; \mathrm{mM}$ citric acid monohydrate
    \item PBS buffer $\mathrm{pH \; (pD)} = 7.4$, $2.7 \; \mathrm{mM}$ \ce{KCl}, $137 \; \mathrm{mM}$ \ce{NaCl}, and $10 \; \mathrm{mM}$ phosphate
\end{itemize}{}
Buffers were prepared in 
\begin{itemize}
    \item \ce{H2O}  (Milli-Q purity with resistivity $\approx 18.2 \; \Omega \cdot \rm{cm}$; Merck Millipore, Molsheim, France)
    \item $100\%$, $80\%$, $60\%$, and $40\%$ of \ce{D2O} in \ce{H2O} corresponding to a molar fraction of \ce{D2O}, $x_{\ce{D2O}} = 1.0, \; 0.8, \; 0.6, \; \& \;0.4$, respectively.
    \item $4.55\%\mathrm{wt.}$, $3.80\%\mathrm{wt.}$, $2.99\%\mathrm{wt.}$ and $2.12\%\mathrm{wt}.$ of glycerol in \ce{H2O} which match the kinematic viscosity, $\nu$, of $100\%$, $80\%$, $60\%$, and $40\%$ \ce{D2O} solutions, respectively.
    \item $10.86\%\mathrm{wt.}$ of glycerol in \ce{H2O} which match the product of the density and viscosity, $\rho \eta$, of $100\%$ \ce{D2O}.
\end{itemize}{}
The physical properties of all these solutions can be found in Table S1 in the SI.

The pH of buffers in \ce{H2O} and glycerol/\ce{H2O} mixtures were determined using a Mettler-Toledo (Ohaio, US) pH meter. The pH (pD) for \ce{D2O}-containing buffers were determined using pH paper, the color of which was compared and matched against the color of a similar pH paper after submerging in the corresponding \ce{H2O}-based buffer. This procedure was followed to make sure that the buffers prepared in \ce{H2O}, \ce{D2O}, \ce{D2O}/\ce{H2O} mixtures, or glycerol/\ce{H2O} mixtures had the same pH (pD). Note that buffer dissociation constants depend on the environment they are in; so the buffer will behave differently in \ce{H2O} than in a glycerol/\ce{H2O} mixture. For this reason, adding equal amounts of the chemicals does not guarantee having the same pH; it is therefore necessary to check and adjust the pH of each solution.

\ce{H2O}-based buffers were either sterlized by autoclaving at $120$ \textcelsius{} for $20 \; \mathrm{minutes}$ or by sterile filtration using $0.22 \; \mathrm{\mu M}$ Stericup-GV Sterile Vacuum filters (Millipore, France); all other buffers were sterile filtered using $0.22 \; \mathrm{\mu M}$ Stericup-GV Sterile Vacuum filters (Millipore, France). Note that the buffers filtered through the $0.22 \; \mathrm{\mu M}$ Stericup-GV Sterile Vacuum filters caused aggregation of the \ce{SiO2} NPs, suggesting that the filter membrane released leachate in the buffer. This problem was addressed by either pre-washing the filters with buffer or autoclaving the buffers instead of filtering them, the later was only done for \ce{H2O}-based buffers.

\subsection*{Gold nanoparticles synthesis}
Gold nanoparticles (Au NPs) were synthesized and surface-modified with acid-terminated poly(ethylene) glycol (PEG) with different molecular weights following the procedure outlined by Perrault and Chen (2009) \cite{Perrault2009}. A seed solution of small Au NPs were made by quickly heating a 100 ml water solution containing 0.01\% (w/v) of \ce{HAuCl4}. Once boiling the solution was kept under vigorous stirring whereupon 3 ml of 1\% (w/v) sodium citrate solution was quickly added. The mixture was allowed to continue to boil for 10 min developing a ruby-red color. To make Au NPs with a diameter of approximately 100 nm, 100 ml water solution containing 0.025\% (w/v) \ce{HAuCl4} was mixed with 0.8 ml of seed solution. The mixture was rapidly stirred at room temperature whereupon $550 \; \mu l$ of 1\% (w/v) sodium citrate solution immediately followed by 2.5 ml of a 30 mM hydroquinone solution. The solution was continuously stirred for another 60 min until its color had stabilized indicating that the reaction was completed. The size of the Au NPs was determined using AFM to be $\approx 70 \; \mathrm{nm}$ in diameter. The Au NP concentration was calculated from the particle size and the amount of gold added to the synthesis assuming that all gold was consumed during the reaction.

Au NPs were surface modified by reaction with acid-terminated PEG-thiols with molecular weight 458.6 Da [(O-(2-Carboxyethyl)-O'-(2-mercaptoethyl)heptaethylene glycol, $>95\%$ purity, Sigma-Aldrich)], 2100 Da [Poly(ethylene glycol) 2-mercaptoethyl ether acetic acid, Sigma-Aldrich, or 5000 Da (alfa-carboxy-beta-mercapto-PEG). Based on the Au NP size and concentration the total surface area of the solution was calculated. Prior to surface modification the Au NPs were concentrated 10 times by centrifugation (1000 g, 20 min) and dilution of the obtained pellet in Milli-Q water. The concentrated Au NPs were mixed with the different thiol-PEG molecules to reach a final concentration of approximately 10 molecules per $\mathrm{nm^2}$ available Au NP surface area. The mixtures were incubated overnight whereupon the PEG-modfied Au NPs were separated from unbound thiol-PEG and salts by four repeated runs through centrifuge filer columns with 300 kDa cut-off (PALL). After the final washing step Au NPs were kept as a concentrated stock solution until use.

\subsection*{POPC vesicles preparation}
POPC was dissolved in chloroform and dried in $50 \; mL$ round flask under vacuum at $60$ \textcelsius{} using a rotavap setup. The dried lipids were left under vacuum overnight to get rid of any residual chloroform. The formed POPC lipid film was then rehydrated in \ce{H2O}-based buffer to a concentration of $1 \; mg \cdot mL^{-1}$, followed by a very brief bath sonication to dissolve any small traces of lipids off the walls of the flask. The POPC solution under went five freeze/thawing cycles. After which, the sample was extruded $31$ times through $30 \; nm$ polycarbonate membranes (Whatman, UK) using a mini-extruder (Avanti, USA). The size of the vesicles were finally determined using dynamic light scattering (DLS) on Zetasizer Nano ZS (Malvern Panalytical, UK).

\subsection*{DNA tethers preparation}
A 45 base pair DNA tether decorated with cholesterol anchors was prepared by mixing the different oligonucleotides (ssA, ssB, ssC, and ssD) as detailed hereafter. $800 \; \mathrm{\mu l}$ of $10 \; \mathrm{\mu M}$ ssA was mixed with ssB (AB), as well as ssC with ssD (CD). The mixtures were vortexed and left to hybridize for at least 30 min at 4 \textcelsius.  Afterwards, the AB and CD segments were mixed together by vortexing and left to hybridize for at least 30 min at 4 \textcelsius. The ABCD dsDNA was diluted in PBS to a final concentration of $0.5 \; \mathrm{\mu M}$.

\subsection*{QCM experiments}
The QCM experiments were conducted in an E4 system (Biolin, Sweden) coupled to GX-274 Autosampler (Gillson, USA) on AT-cut 5 MHz quartz sensors, coated with \ce{SiO2}. The sensors were either used without modification or after silanization with amine-terminated silanes. Bare sensors were cleaned by bath sonication for 15 minutes in 2\%wt.\ SDS solution, followed by rinsing with Milli-Q water, then drying under a \ce{N2} flow, and finally treating with \ce{O2} plasma for 3 minutes directly before use. The temperature and flow rate was kept constant all through the QCM experiments at 25 \textcelsius{} and $50 \; \mathrm{\mu l \cdot min^{-1}}$, respectively. At the end of the Au and \ce{SiO2} NPs adsorption experiments, the solution in contact of the sensor was exchanged to ethanol $\geq 99.5 \%$, then the sensors were dried \textit{in-situ} under \ce{N2} flow and at temperature of 60 \textcelsius{}; finally the temperature were lowered down again to 25 \textcelsius{}. This procedure was followed to dry the sensors for determining the response after drying (in air) and for further analysis using AFM and SEM while minimizing any drying effects, such as aggregation of the NPs on the surface. A step-by-step experimental and calculation procedure are presented in Sections S3 and S4.

\subsection*{Silanization of \ce{SiO2} sensors}
The sensors were first cleaned by: bath sonication for 15 minutes in 2 \%wt.\ SDS solution, rinsing with MiliQ water, treating with \ce{O2} plasma for 3 minutes, adding few drops of $4 \; \mathrm{M}$ \ce{H2SO4} on the surface of the sensor for $30 \; \mathrm{minutes}$, rinsing thoroughly under flowing MilliQ water, rinsing thoroughly using anhydrous methanol, and finally drying under \ce{N2} flow. The sensors were then incubated in $7.5 \; \%\mathrm{v/v}$ 3-(Ethoxydimethylsilyl)propylamine in anhydrous methanol for at least $30 \; \mathrm{minutes}$. Afterwards the sensors were rinsed thoroughly with anhydrous methanol to avoid precipitation of the silanes which happens in the presence of water or ethanol. Finally the sensors are rinsed with MilliQ water and mounted in the instrument for use.

\subsection*{Atomic Force Microscopy (AFM)}
A Bruker Dimension 3100 scanning probe microcopy system was used for \textit{ex-situ} atomic force imaging of the surface of dried QCM sensors. The height (z-distance) of both \ce{SiO2} and Au NPs were used to determine their size. A detailed experimental and computation procedure are presented in Section S5.

\subsection*{Scanning Electron Microscopy (SEM)}
Scanning electron microscopy (Zeiss Ultra 55 FEG, in-lens mode, ETH=2 kV) was used for general \textit{ex-situ} characterization of the dried QCM sensor surfaces and for determination of NP surface coverage. Images were acquired at a location as close as possible to the center of the sensor; the area of the sensor which contributes the most to the QCM signal. We also tried to select an imaging area that did not show any visible drying effects, such as clear aggregation of the NPs. Based on the acquired images NPs were counted using ImageJ; the total number of the NPs was then divided by the area of the image to obtain the number of $\mathrm{NPs \; \mathrm{per \; \mu m^2}}$. Features much larger or much smaller than the size of the NPs were not counted towards the total number of the NPs. Features which showed a double layer of NPs was double counted. Examples of these different cases, as well as a detailed experimental and computation procedure are presented in Section S5 in the SI.

\subsection*{Surface resonance plasmon (SPR) experiments}
The SPR experiments were conducted using a dual wavelength, $670$ and $785 \; \mathrm{nm}$, multi-parametric SPR Navi\texttrademark  420A (BioNavis, Finland) on silica-coated sensors. The sensors were cleaned by bath sonication for 15 minutes in 2\%wt.\ SDS solution, followed by rinsing with Milli-Q water, then drying under a \ce{N2} flow, and finally treating in a UV/Ozone chamber for 45 minutes directly before use. The temperature and flow rate were kept constant at 25 \textcelsius{} and $7 \; \mathrm{\mu l \cdot min^{-1}}$, respectively. 

\newpage

\section*{Supporting Information}
Detailed theoretical derivation; calculation of the physical properties of different \ce{D2O}/\ce{H2O} and glycerol/\ce{H2O} mixtures; detailed methods and experimental procedure to measure the response in different liquids and for independent mass determination using measurements in air and SEM; experimental results for measurements with reference to air, polymers adsorption to APDMES-coated sensors, adsorption of carboxylated and PEGylated Au NPs, response at different overtones and \ce{D2O} concentrations, lipid vesicles characterization and tethering, SPR calculations and results.

\section*{Acknowledgments}
The authors thank Bengt Kasemo and Michael Rodal for valuable discussions and Andreas Dahlin and John Andersson for support with SPR analysis. This work was performed in part at the Material Analysis Laboratory (CMAL) and infrastructure for nanofabrication (MC2) at Chalmers. The authors thank the Knut and Alice Wallenberg Foundation and the Swedish Research Council (2018-04900) for funding.

\section*{Author contributions}
A.A. and F.H. conceived the idea. V.P.Z. developed the theoretical formalism. A.A., B.A., and A.L designed and performed the experiments. A.A., V.P.Z. and F.H. wrote the paper with input from all authors. 

\section*{Notes}
The authors declare no competing interests. 

\newpage

\bibliographystyle{unsrt}
\bibliography{library.bib}

\end{document}